\documentclass[useAMS,usenatbib]{mn2e}

\usepackage{times}
\usepackage{graphicx}
\usepackage{subfigure}
\usepackage{supertabular}
\usepackage{amssymb}
\usepackage{amsbsy}
\usepackage{amsmath}
\usepackage{placeins}
\usepackage{comment}
\usepackage{pifont}
\usepackage{color}
\title[Recombination studies of G333.6$-$0.2]{Radio and infrared recombination
studies of the southern massive star-forming region G333.6{\boldmath $-$}0.2}
\author[T. Fujiyoshi et al.]{Takuya Fujiyoshi,$^{1,2}$\thanks{E-mail:
tak@subaru.naoj.org (TF);
csmith@eostech.com (CHS);
jcaswell@atnf.csiro.au (JLC);
tjtm@astro.livjm.ac.uk (TJTM);
sll@ast.leeds.ac.uk (SLL);
dka@star.herts.ac.uk (DKA);
pfr@astro.ox.ac.uk (PFR)}
Craig H. Smith,$^{3}$\footnotemark[1]
James L. Caswell,$^{4}$\footnotemark[1]
Toby J. T. Moore,$^{5}$\footnotemark[1]
\newauthor
Stuart L. Lumsden,$^{6}$\footnotemark[1]
David K. Aitken$^{7}$\footnotemark[1]
and Patrick F. Roche$^{8}$\footnotemark[1] \\
$^{1}$Subaru Telescope, National Astronomical Observatory of Japan,
National Institutes of Natural Sciences, 650 North A'ohoku Place,
Hilo, HI 96720, USA \\
$^{2}$Department of Physics, University College, University of New South
Wales, Australian Defence Force Academy, Canberra ACT 2600, Australia \\
$^{3}$Electro Optic Systems Pty Limited, 55A Monaro Street, Queanbeyan
NSW 2620, Australia \\
$^{4}$Australia Telescope National Facility, CSIRO, P.O.\ Box 76, Epping
NSW 2121, Australia \\
$^{5}$Astrophysics Research Institute, Liverpool John Moores University,
Twelve Quays House, Egerton Wharf, Birkenhead CH41 1LD \\
$^{6}$Department of Physics and Astronomy, University of Leeds, Leeds LS2
9JT \\
$^{7}$Department of Physical Sciences, University of Hertfordshire,
Hatfield, Herts AL10 9AB \\
$^{8}$Astrophysics, Department of Physics, Oxford University, Keble
Road, Oxford OX1 3RH
}

\begin{document}

\date{To be published in MNRAS}

\pagerange{\pageref{firstpage}--\pageref{lastpage}} \pubyear{2006}

\maketitle

\label{firstpage}

\begin{abstract}
We present high spatial resolution radio and near-infrared hydrogen
recombination line observations of the southern massive star-forming
region G333.6$-$0.2. The 3.4-cm continuum peak is found slightly
offset from the infrared source. The H90$\alpha$ spectra show for the
first time a double peak profile at some positions. The complex
velocity structure may be accounted for by champagne outflows, which
may also explain the offset between the radio and infrared sources.
The 2.17-$\umu$m Br$\gamma$ image and H90$\alpha$ map are combined to
construct an extinction map which shows a trend probably set by the
blister nature of the H\,{\sc ii} region. The total number of Lyman
continuum photons in the central 50-arcsec is estimated to be equivalent
to that emitted by up to 19 O7V stars.
\end{abstract}

\begin{keywords}
H\,{\sc ii} regions -- ISM : individual : G333.6$-$0.2 -- infrared : ISM --
radio continuum : ISM -- radio lines : ISM
\end{keywords}

\section{Introduction}

G333.6$-$0.2 is a southern compact H\,{\sc ii} region which was initially
detected as one of the brightest radio sources in a 5-GHz Galactic plane
survey (Goss \& Shaver 1970). Radio recombination lines (RRLs) H90$\alpha$
and He90$\alpha$ were observed by McGee, Newton \& Batchelor (1975), and
McGee \& Newton (1981) examined H76$\alpha$ and He76$\alpha$ RRLs.
Both papers reported a Local Thermal Equilibrium (LTE) electron
temperature ($T_{e}^{*}$) of slightly less than 7000~K and a
ratio of singly ionised helium to singly ionised hydrogen
[$Y^{+}=\frac{N({\rm He}^{+})}{N({\rm H}^{+})}$] of less than the solar
value (i.e., $<10$~per~cent). The H\,{\sc ii} region is almost totally
obscured in the visible ($A_{V}=18.5$, Landini et al.\ 1984) but is bright
at infrared (IR) wavelengths, and has been studied intensively in
this region of the spectrum (e.g.\ Aitken, Griffiths \& Jones 1977;
Hyland et al.\ 1980; Fujiyoshi et al.\ 1998, 2001, 2005, hereafter
Papers~1, 2, and 3, respectively). In Paper~1, it was found that the
H\,{\sc ii} region must be excited by a cluster of O and B stars, rather
than by a single star.  Indeed, near-IR (NIR) broadband imaging revealed
multiple point sources in G333.6$-$0.2 (Paper~3). Point sources with
significant $H-K$ colour excess were found preferentially in extended
emission regions and it was suggested that these red objects were
physically associated with the H\,{\sc ii} region. The powerful nature of
this nebula was unveiled by mid-IR (MIR) imaging polarimetry (Paper~2)
which showed bent magnetic field lines possibly shaped by the
expansion of the ionised gas.

At radio wavelengths the most intense emission from H\,{\sc ii} regions
is caused by interactions between unbound charged particles, better known
as free-free radiation or {\em bremsstrahlung} (braking radiation).
Since the particles are free and their energy states are not quantised
the radiation resulting from changes in their kinetic energy is continuous.
Free-free emission is dominated by interactions of free electrons with the
most abundant element, singly ionised hydrogen. Observations of the
radio free-free emission (where extinction is usually negligible) can be
used to predict the hydrogen recombination line fluxes at IR wavelengths,
which in turn can be compared to the observed values to estimate extinction
at the line wavelengths. 

In this paper, we present radio synthesis observations at 3.4~cm, and the
2.17-$\umu$m Br$\gamma$ hydrogen recombination line imaging of
G333.6$-$0.2. High spatial resolution data provided by radio
interferometry can be compared directly with those at NIR wavelengths
to infer spatial variation of extinction across the face of the H\,{\sc ii}
region. The H90$\alpha$ RRL, observed both at high spatial and spectral
resolution, is used as a probe of the physical conditions in the emitting
gas.

\section[]{Observations and data reduction}

\subsection{Radio observations}

The radio observations were made on 1996 October 8 at the
Australia Telescope Compact Array (ATCA) in Narrabri, Australia.
The ATCA is a synthesis telescope which consists of six 22-m antennae
on a 6-km east-west track. One set of measurements was taken with
the baseline configuration 6A. The longest baseline of 6~km was chosen
to give the final map the highest possible spatial resolution available
with the ATCA.

G333.6$-$0.2 was observed at a frequency centred between the
H90$\alpha$ (rest frequency 8.872~GHz) and He90$\alpha$ (8.876~GHz) RRLs.
Each of two orthogonal polarisations was recorded, using a correlator
configuration of 512 channels covering a 16-MHz bandwidth.
The primary calibrator (1934$-$638; 2.8~Jy at 8.8~GHz with an uncertainty of
up to a few~per~cent) was observed for 20~minutes during the run for the
bandpass correction and the flux calibration. A secondary calibrator
(1619$-$680) was observed every 20~minutes for 5~minutes throughout the
observation period for position and phase calibration.

Data reduction was carried out using the software package MIRIAD. The
continuum and H90$\alpha$ RRL maps were constructed using natural weighting
in the conventional Fast Fourier Transformation. The maps were then
deconvolved using the CLEAN routine in MIRIAD. Spectral cubes were also
made following the same method; however, in order to improve the
signal-to-noise ratio ($S/N$), 4 consecutive channels were binned together
resulting in an effective velocity resolution of $\sim4$~km~s$^{-1}$.
The cleaned maps were finally restored to intensity maps with a
1.6-arcsec diameter circular beam, which therefore is the effective
spatial resolution of the maps presented here.

\subsubsection{Missing flux from extended structures}
\label{missing.flux}

McGee et al.\ (1975) observed G333.6$-$0.2 as part of their H90$\alpha$
RRL survey of bright southern radio sources. Using a 2.5-arcmin beam
they measured the 3.4-cm continuum flux density of $80\pm17$~Jy.
However, in our continuum map the integrated total flux density in an
area $\sim2\times2$~arcmin$^{2}$ was found to be only $\sim13$~Jy.

Lack of short baselines in interferometry leads to the loss of
sensitivity to extended structures. The ATCA 6A configuration has a
shortest baseline of 337~m, and thus at 3.4~cm is not responsive to
structure smoothly distributed on a scale of $\sim30$~arcsec or larger.
The target is a compact H\,{\sc ii} region whose IR counterpart has
a core-halo geometry (Aitken et al.\ 1977) in which most flux is
concentrated in the central region.
We have assumed that the missing
extended emission underlying the compact H\,{\sc ii} region takes the
shape of a 2-D Gaussian. Its peak value
and full widths at half maximum (FWHMs) are adjusted so that it would
contain the missing flux and satisfy the level of extinction measured
by Landini et al.\ (1984) (see Section~\ref{extmap.sect} for details).

\subsection{NIR observations}

The NIR observations of G333.6$-$0.2 were carried out on 1995 August 11.
All measurements were obtained using the IR imaging spectrometer
IRIS at the $f/36$ Cassegrain focus of the 3.9-m Anglo-Australian Telescope
(AAT), Siding Spring, Australia. The detector element of IRIS is a Rockwell
$128\times128$ HgCdTe NICMOS2 array, which is housed in a closed-cycle
compressed helium refrigeration unit. In the imaging mode, wavelength
selection is provided by discrete filters such as `standard' NIR broadband
($J, H, K$). A more complete description of the instrument can be found in
Allen et al.\ (1993).

Table~\ref{tab1} summarises the NIR observations of G333.6$-$0.2.
INTERM mode was used to give a pixel size of
$\sim0.25\times0.25$~arcsec$^{2}$. The object was dithered in
the  instrument's field of view, which produced the final image size of
$\sim50\times50$~arcsec$^{2}$. The 2.25-$\umu$m H$_{2}$(2$-$1) S(1) image
was used for subtracting the continuum from the 2.17-$\umu$m Br$\gamma$
image. Moneti \& Moorwood (1989) found that G333.6$-$0.2 showed only a weak
H$_{2}$(1$-$0) S(1) emission in the central region
[$I_{{\rm Br}\gamma}=3.5\times10^{-18}$~W~cm$^{-2}$ and
$I_{(1-0)}=1.3\times10^{-20}$~W~cm$^{-2}$, both measured in a
$6\times6$-arcsec$^{2}$ beam].
In conditions characteristic of H\,{\sc ii} regions, the (1$-$0) mode is
always much stronger than the (2$-$1) (e.g.\ Burton 1992). We therefore
assumed that the (2$-$1) image contained only the continuum emission. The
calibration was achieved by observing the standard star SA94$-$251
($K=8.305$, Carter \& Meadows 1995). Seeing on the night was quite good
and observations of standard stars showed FWHMs of about 0.6~arcsec.

\begin{table}
  \caption{Summary of NIR observations of G333.6$-$0.2. \label{tab1}}
  \begin{tabular}{@{}lllll@{}}
  \hline
  Date        & Filter              & $\lambda_{0}$ & $\Delta\lambda$ & Integration \\
              &                     & ($\umu$m)     &                 & (sec/position) \\
  \hline
  11 Aug 1995 & Br$\gamma$          & 2.17          & 1~\%            & $2\times10.0$ \\
              & H$_{2}$(2$-$1) S(1) & 2.25          & 1~\%            & $1\times10.0$ \\
 \hline
 \end{tabular}
\end{table}

\section{Results and discussions}

\subsection{Radio data}

\subsubsection{The 3.4-cm continuum}

\begin{figure}
\includegraphics[width=84mm]{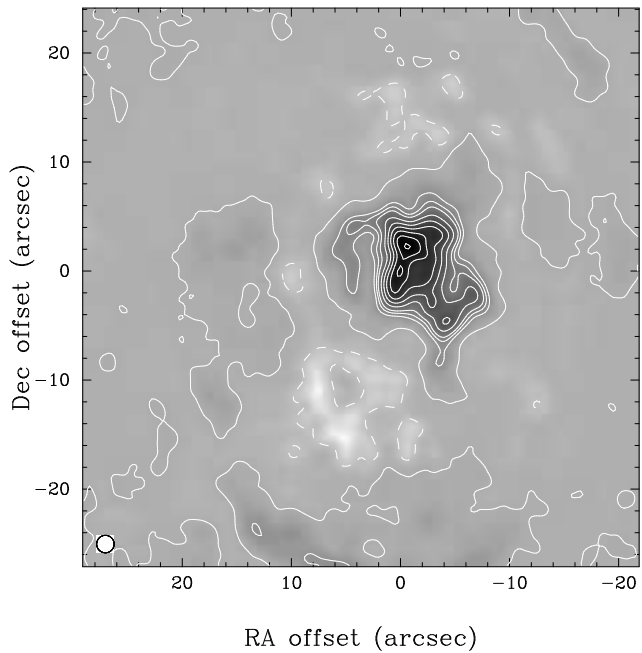}
\caption{The 3.4-cm continuum grey-scale map of G333.6$-$0.2. Contours
are drawn at $-$0.050 (broken line), 0.00050, 0.10, 0.15, 0.20, 0.30, 0.40,
0.50, 0.60, 0.70, and 0.79 Jy~beam$^{-1}$. Axis offsets are from the MIR
main peak position
[${\rm RA}=16^{{\rm h}}22^{{\rm m}}09.\!\!^{{\rm s}}6$
and ${\rm Dec}=-50^{\circ}06'00''$ (J2000)]. The circular 1.6-arcsec
diameter synthesised beam is also shown at the bottom left hand corner.}
\label{rad.cont}
\end{figure}

Figure~\ref{rad.cont} shows the 3.4-cm continuum map. The overall
similarity to the IR images (cf.\ Figure~\ref{brg.after}. See also
Papers~1 \& 3) is quite remarkable. It is dominated by a north-south
elongated main peak in the centre with steeper gradient to the east.
Hyland et al.\ (1980) suggested that this steep gradient
may be caused by extinction. Indeed, the extinction map (see
Section~\ref{extmap.sect}) shows a slight increase of extinction in
this direction. An arm-like extension, which was described as the
secondary peak in Papers~1 \& 3, is seen about 5~arcsec to the east
of the main peak. The gap between the main and secondary peaks is also
present. As the gap is still present at 3.4~cm, it is likely that the
main and secondary peaks are separate ionised regions and the gap
represents a narrow non-ionized region in between, rather than an
apparent structure simply caused by a heavy foreground extinction.
It was suggested in Paper~3 that both main and secondary peaks house
exciting sources.

The most significant difference between the IR and radio continuum
images is that the radio continuum main peak is found offset slightly
($\sim2$~arcsec) from the MIR main peak position. The coordinates of
the radio main peak are
${\rm RA}=16^{{\rm h}}22^{{\rm m}}09.\!\!^{{\rm s}}5$ and
${\rm Dec}=-50^{\circ}05'58''$ (J2000).
The size of the source, determined by fitting a 2-D Gaussian profile, is
4.9~arcsec in RA and 6.9~arcsec in Dec, or about 0.086~pc
$\sim1.8\times10^{4}$~au at 3.0~kpc (an assumed distance to G333.6$-$0.2,
Colgan et al.\ 1993). There is a ridge running to the south (and
slightly east) from the main continuum peak and in fact the MIR main peak
position coincides with the peak on this ridge.

\subsubsection{The H90$\alpha$ RRL}

\begin{figure}
\includegraphics[width=84mm]{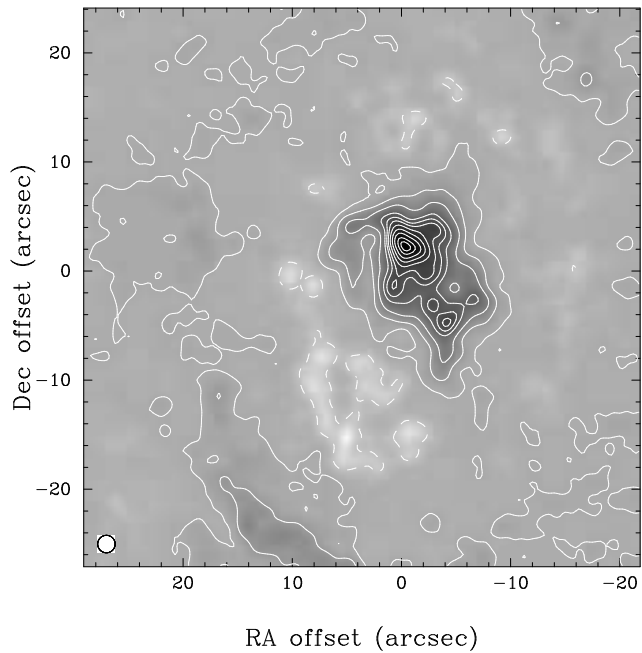}
\caption{The H90$\alpha$ grey-scale map of G333.6$-$0.2. Contours are
drawn at
$-$0.25 (broken line), 0.025, 0.25, 0.65, 1.3, 1.7, 2.1, 2.5, 3.0, 3.5, 4.0,
4.3 Jy$\cdot$km~s$^{-1}$~beam$^{-1}$. The coordinates of the origin are
as in Figure~\ref{rad.cont}.}
\label{rad.line}
\end{figure}

The H90$\alpha$ RRL map is presented in Figure~\ref{rad.line}.
It shows the spatial variation of the integral over velocity of the
recombination line. The distribution of the RRL emission is similar
to that of the 3.4-cm continuum. The size of the H90$\alpha$ peak
emission, which is coincident with the peak of the 3.4-cm continuum
emission, measured again by fitting a 2-D Gaussian profile, is
4.1~arcsec in RA and 5.7~arcsec in Dec, which is only slightly smaller
than the size of the continuum peak. There is also a ridge running to
the south of the main peak; however, the peak on the ridge is slightly
more south-east (by about 1~arcsec south and 0.5~arcsec east) than the
ridge peak position in the 3.4-cm continuum map.

\begin{figure}
\includegraphics[angle=270,width=84mm]{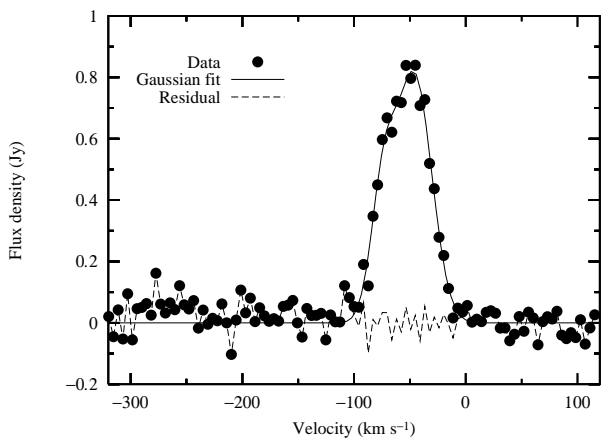}
\caption{The total H90$\alpha$ spectrum in the central
$\sim2\times2$~arcmin$^{2}$. It shows the data points (dots),
a multiple (double) Gaussian fit to the data (solid line),
and the residual (i.e., the difference between the actual data and
the fit; dashed line).}
\label{whole.spec}
\end{figure}

Figure~\ref{whole.spec} shows the total H90$\alpha$ RRL spectrum in the
central $\sim2\times2$~arcmin$^{2}$. Attempts were made to fit
either a single or a double Gaussian profile to the spectrum and it was
found that a combination of two Gaussian functions gave a better fit
(the sum of fitting residuals was smaller). As it will be shown later,
G333.6$-$0.2 exhibits a complex velocity structure with some positions
displaying a double peak spectrum. Such a structure has never been
observed with large-beam, single-dish radio studies of the H\,{\sc ii}
region and this emphasises the importance of high spatial resolution
investigations.

The parameters derived from the fit are summarised in
Table~\ref{whole.tab}. All velocities refer to the Local Standard of
Rest (LSR) frame. The weak broad feature centred at around
$-270$~km~s$^{-1}$ is the H113$\beta$ RRL. Additional weak emission
may be present at about $-150$~km~s$^{-1}$ and could be the
He90$\alpha$ RRL, although it would normally be expected at a bluer
velocity [$\sim-175$~km~s$^{-1}$, cf.\ Figure~1(o), McGee et al.\ (1975)].
It should be noted, however, that both lines presumably also exhibit
double peak spectra and therefore are likely to be blended.
No attempt was made to fit parameters to these RRLs.

\begin{table}
\caption{The Gaussian fit parameters to the total H90$\alpha$ spectrum in
the central $\sim2\times2$~arcmin$^{2}$ (see Figure~\ref{whole.spec}).}
\label{whole.tab}
\begin{tabular}{llll}
\hline
$S_{l}$          & $V_{{\rm LSR}}$ & $\Delta V_{G}$ & Reference \\
(mJy)            & (km~s$^{-1}$)   & (km~s$^{-1}$)  & \\
\hline
$433\pm117$      & $-73.4\pm3.1$   & $26.3\pm4.5$   & This study \\
$798\pm53$       & $-45.6\pm2.8$   & $36.1\pm4.2$   & This study \\
                 & $-47.1\pm0.2$   & $43.9\pm0.3$   & McGee et al. (1975) \\
\hline
\end{tabular}
\end{table}

Table~\ref{whole.tab} also lists the single-dish H90$\alpha$ measurements
of McGee et al.\ (1975). The overall velocity of the ionized gas in
G333.6$-$0.2 was found by three independent single-dish observations
at different frequencies to be $\sim-46$~km~s$^{-1}$. Apart from McGee
et al.\ (1975), who found its velocity to be $-47.1$~km~s$^{-1}$, McGee
\& Newton (1981) determined $-45.3$~km~s$^{-1}$ from the 14.7-GHz
H76$\alpha$ RRL observation with a 2.3-arcmin beam, and Caswell \&
Haynes (1987) obtained $-46$~km~s$^{-1}$ by measuring the 5-GHz
H109$\alpha$ and H110$\alpha$ RRLs with a $\sim4$-arcmin beam.
The redder ($\sim-46$~km~s$^{-1}$) of the two velocity components
we found above is therefore likely to be associated with emission from
extended structures, most of which escaped detection in the current
study (see Section~\ref{missing.flux}). The bluer component at
$\sim-70$~km~s$^{-1}$, detected for the first time with high spatial
resolution observations presented here, must then be originating from
compact structures.

\begin{figure*}
\includegraphics[width=125mm]{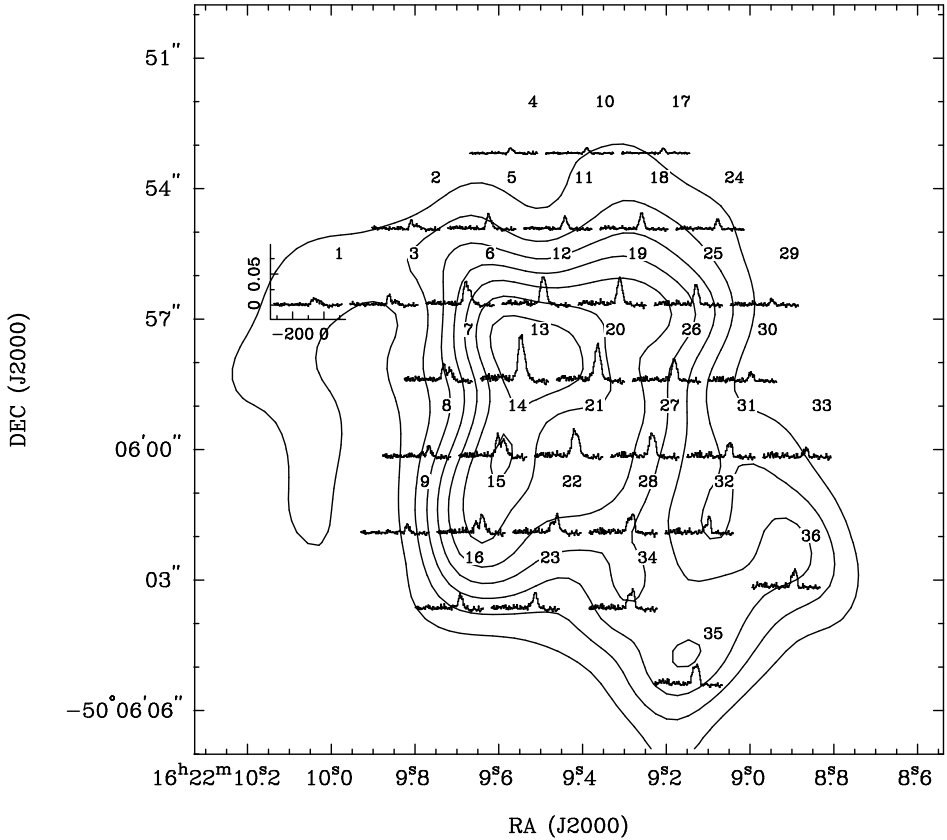}
\caption{A grid of average spectra in $1.75\times1.75$-arcsec$^{2}$ boxes
superimposed on the 3.4-cm continuum map of G333.6$-$0.2. Contours are
drawn from 0.1~Jy~beam$^{-1}$ in an 0.1~Jy~beam$^{-1}$ interval.}
\label{grid.spec}
\end{figure*}

\begin{figure*}
\centering
\includegraphics[width=150mm]{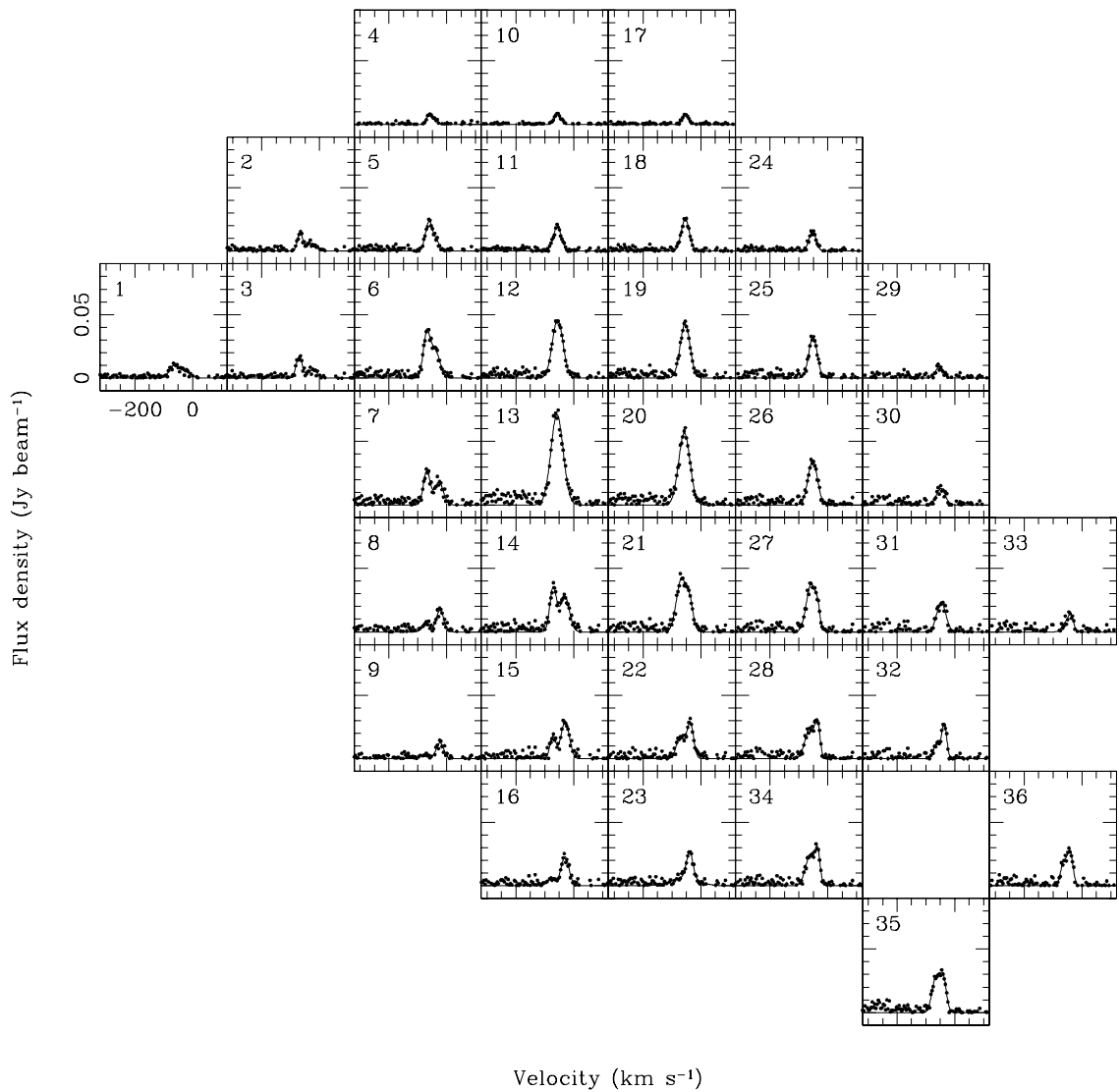}%}%
\caption{The H90$\alpha$ RRL spectra averaged in
$1.75\times1.75$-arcsec$^{2}$ boxes (see Figure~\ref{grid.spec}). Each
window shows data points (dots), a Gaussian  profile fit to the data
(solid line), and the number at the top left hand corner refers to the
position number in Figure~\ref{grid.spec}.}
\label{rad.spec}% label for figure
\end{figure*}

Figure~\ref{grid.spec} shows average spectra in
$1.75\times1.75$-arcsec$^{2}$ boxes in the central region superimposed
on the 3.4-cm continuum map. The boxes are positioned so that they would
be centred on the main peak and the peak on the ridge. Positions~34, 35,
and 36 are centred on each of the three radio peaks in the south-west
extension (see Figures~\ref{rad.cont} \& \ref{rad.line}).
Figure~\ref{rad.spec} shows the H90$\alpha$ RRL spectra fitted with
Gaussian profiles. Again, we determined whether each spectrum contains a
single or a double line by fitting either a single or a double Gaussian
profile and selecting the one that resulted in a smaller sum of residuals.
Parameters found by the fit are summarised in Table~\ref{grid.num}.

\begin{table*}
\begin{minipage}{160mm}
\caption{The Gaussian fit parameters to the H90$\alpha$ RRL from the
grid of $1.75\times1.75$-arcsec$^{2}$ boxes (see Figure~\ref{rad.spec}).
The position numbers refer
to those indicated in Figure~\ref{grid.spec}.
$S_{l}$ is in mJy~beam$^{-1}$ and $n_{e}$ in $\times10^{5}$~cm$^{-3}$.
}\label{grid.num}
\begin{tabular}{rrrrrrrrrrrrrrr}
\hline
No.      & $S_{l}$    & err   & $V_{{\rm LSR}}$ & err  & $\Delta V_{G}$ & err & $T_{e}^{*}$ & $T_{b}$ & $\tau_{c}$ & $T_{e}$ & err   & $n_{e}$ & $b_{n}$  & $\beta_{n}$ \\
         &            &       & (km/s)          &      & (km/s)         &     & (K)         & (K)     &            & (K)     &       &         &          & \\
\hline
1        & 9.3        & 4.1   & $-$65.7         & 7.4  & 35.8           & 9.8 & 4861        &  856    & 0.19       & 5000    &  4000 & 3.00    & 0.997977 & $-$0.372641 \\
         & 5.4        & 2.0   & $-$29.4         & 18.8 & 46.1           & 30.1& & & & & & & & \\
2        & 14.0       & 1.2   & $-$68.1         & 0.9  & 20.3           & 2.2 &  3166       &  488    & 0.17       & 3200    &   800 & 1.65    & 0.997272 & $-$0.184034 \\
         & 6.1        & 0.7   & $-$31.4         & 3.7  & 42.3           & 9.1 & & & & & & & & \\
3        & 16.9       & 1.2   & $-$70.8         & 1.0  & 24.1           & 2.4 &  4477       &  883    & 0.22       & 4500    &  1200 & 4.58    & 0.998264 & $-$0.069765 \\
         & 6.3        & 1.0   & $-$24.9         & 3.4  & 40.1           & 9.6 & & & & & & & & \\
4        & 8.0        & 0.9   & $-$59.4         & 2.5  & 21.1           & 4.2 &  3093       &  209    & 0.07       & 3100    &  1600 & 1.65    & 0.997268 & $-$0.149526 \\
         & 3.6        & 1.0   & $-$38.7         & 5.0  & 19.5           & 9.3 & & & & & & & & \\
5        & 23.7       & 1.3   & $-$60.2         & 1.9  & 29.9           & 4.1 &  3834       &  935    & 0.27       & 3900    &  2100 & 2.51    & 0.997812 & $-$0.160766 \\
         & 6.6        & 2.7   & $-$34.2         & 4.5  & 19.5           & 8.4 & & & & & & & & \\
6        & 30.7       & 9.4   & $-$69.7         & 2.7  & 27.7           & 4.2 &  4350       & 2269    & 0.59       & 5100    &  2100 & 1.90    & 0.997535 & $-$0.691416 \\
         & 21.8       & 3.8   & $-$41.3         & 8.2  & 41.5           & 11.5& & & & & & & & \\
7        & 26.6       & 1.9   & $-$68.7         & 1.3  & 27.1           & 3.0 &  4349       & 1773    & 0.47       & 4700    &   900 & 2.44    & 0.997793 & $-$0.404504 \\
         & 18.1       & 1.7   & $-$26.6         & 2.2  & 36.0           & 5.6 & & & & & & & & \\
8        & 7.2        & 1.5   & $-$71.7         & 3.3  & 28.9           & 8.6 &  4980       & 1101    & 0.25       & 5000    &  1700 & 7.71    & 0.998473 & $-$0.049835 \\
         & 17.1       & 1.5   & $-$23.0         & 1.3  & 30.5           & 3.4 & & & & & & & & \\
9        & 3.2        & 1.1   & $-$77.5         & 7.1  & 35.9           & 20.5&  4780       &  718    & 0.16       & 4800    &  2900 & 5.66    & 0.998362 & $-$0.078471 \\
         & 12.6       & 1.2   & $-$22.8         & 1.5  & 30.3           & 4.3 & & & & & & & & \\
10       & 8.6        & 0.3   & $-$56.8         & 0.5  & 24.0           & 1.4 &  6063       &  422    & 0.07       & 6200    &  2300 & 2.51    & 0.997843 & $-$0.800377 \\
         & 2.0        & 0.5   & $-$37.5         & 1.1  &  8.7           & 2.7 & & & & & & & & \\
11       & 19.0       & 0.9   & $-$56.5         & 0.7  & 33.6           & 1.8 &  4279       &  810    & 0.21       & 4300    &   300 & 3.95    & 0.998179 & $-$0.070785 \\
12       & 45.9       & 1.4   & $-$55.6         & 0.7  & 43.6           & 1.6 &  4761       & 2869    & 0.70       & 5700    &   200 & 2.51    & 0.997835 & $-$0.664589 \\
13       & 71.3       & 2.2   & $-$57.6         & 0.7  & 49.9           & 1.8 &  4169       & 4381    & 1.09       & 6600    &   300 & 1.62    & 0.997404 & $-$1.284620 \\
14       & 34.6       & 1.8   & $-$71.1         & 1.0  & 27.0           & 2.2 &  6011       & 3716    & 0.68       & 7500    &  1000 & 4.09    & 0.998172 & $-$0.853680 \\
         & 27.7       & 1.4   & $-$30.9         & 1.5  & 37.9           & 4.0 & & & & & & & & \\
15       & 17.5       & 1.9   & $-$72.0         & 1.5  & 23.9           & 3.7 &  7540       & 3461    & 0.47       & 9200    &  1700 & 6.18    & 0.998316 & $-$1.098160 \\
         & 29.5       & 1.7   & $-$29.4         & 1.1  & 34.0           & 3.0 & & & & & & & & \\
16       & 5.7        & 1.2   & $-$78.5         & 3.5  & 30.4           & 9.7 &  5914       & 1654    & 0.29       & 6600    &  2200 & 2.37    & 0.997801 & $-$0.949274 \\
         & 22.7       & 1.1   & $-$30.0         & 0.8  & 32.0           & 2.2 & & & & & & & & \\
17       & 7.1        & 6.7   & $-$56.3         & 11.2 & 18.2           & 11.8&  6513       &  390    & 0.06       & 6600    & 16800 & 4.85    & 0.998259 & $-$0.561682 \\
         & 3.8        & 8.1   & $-$42.8         & 13.6 & 15.9           & 26.4& & & & & & & & \\
18       & 25.2       & 1.0   & $-$53.4         & 0.6  & 32.9           & 1.5 &  4965       & 1246    & 0.29       & 5000    &   300 & 7.14    & 0.998447 & $-$0.066755 \\
19       & 42.7       & 1.5   & $-$54.0         & 0.7  & 39.8           & 1.7 &  5628       & 2958    & 0.62       & 6400    &   300 & 4.76    & 0.998253 & $-$0.520325 \\
20       & 56.7       & 1.7   & $-$56.7         & 0.7  & 46.2           & 1.6 &  4973       & 3953    & 0.96       & 6400    &   300 & 3.19    & 0.998028 & $-$0.705483 \\
21       & 41.3       & 3.1   & $-$66.8         & 3.8  & 40.7           & 5.7 &  5157       & 3470    & 0.68       & 7000    &  3000 & 1.91    & 0.997606 & $-$1.238900 \\
         & 19.2       & 7.9   & $-$37.3         & 4.1  & 27.3           & 6.5 & & & & & & & & \\
22       & 17.1       & 1.3   & $-$71.3         & 2.1  & 30.8           & 4.7 &  7500       & 3194    & 0.45       & 8800    &  1500 & 8.80    & 0.998410 & $-$0.903176 \\
         & 28.8       & 1.4   & $-$36.7         & 1.2  & 27.5           & 2.4 & & & & & & & & \\
23       & 6.5        & 1.1   & $-$55.5         & 6.0  & 104.7          & 14.2&  5279       & 1972    & 0.44       & 5500    &  1200 & 6.07    & 0.998373 & $-$0.224923 \\
         & 20.8       & 1.4   & $-$38.0         & 0.7  & 26.1           & 2.2 & & & & & & & & \\
24       & 15.3       & 0.9   & $-$52.9         & 0.8  & 29.6           & 2.1 &  5104       &  703    & 0.15       & 5200    &   400 & 3.71    & 0.998128 & $-$0.324548 \\
25       & 32.7       & 1.4   & $-$51.3         & 0.7  & 34.0           & 1.8 &  6055       & 2101    & 0.40       & 6400    &   400 & 8.99    & 0.998473 & $-$0.338987 \\
26       & 34.8       & 1.5   & $-$52.3         & 0.8  & 38.4           & 1.9 &  6219       & 2605    & 0.43       & 7500    &   400 & 2.54    & 0.997875 & $-$1.137830 \\
27       & 37.7       & 1.5   & $-$57.2         & 2.1  & 40.0           & 3.5 &  5476       & 2824    & 0.67       & 5800    &  3200 & 8.62    & 0.998480 & $-$0.209978 \\
         & 10.1       & 4.9   & $-$34.6         & 1.7  & 16.6           & 6.7 & & & & & & & & \\
28       & 23.6       & 1.3   & $-$62.6         & 3.0  & 39.3           & 5.6 &  6043       & 2692    & 0.51       & 6700    &  1500 & 6.30    & 0.998360 & $-$0.497484 \\
         & 24.8       & 4.2   & $-$35.6         & 1.0  & 20.2           & 2.6 & & & & & & & & \\
29       & 8.9        & 1.2   & $-$54.1         & 2.0  & 30.0           & 4.9 &  5582       &  460    & 0.08       & 5900    &  1100 & 0.96    & 0.996526 & $-$1.712310 \\
30       & 12.9       & 1.5   & $-$48.5         & 2.0  & 35.7           & 4.7 &  5099       &  716    & 0.15       & 5200    &   800 & 3.64    & 0.998115 & $-$0.332842 \\
31       & 23.2       & 1.7   & $-$47.0         & 1.4  & 37.5           & 3.2 &  4571       & 1190    & 0.23       & 5900    &   600 & 0.48    & 0.994688 & $-$3.089530 \\
32       & 11.3       & 1.4   & $-$61.7         & 5.0  & 29.3           & 11.6&  5369       & 1433    & 0.29       & 5700    &  2200 & 3.31    & 0.998051 & $-$0.505839 \\
         & 26.7       & 3.5   & $-$36.4         & 1.4  & 20.1           & 2.7 & & & & & & & & \\
33       & 6.4        & 9.1   & $-$58.4         & 33.8 & 26.1           & 44.2&  4055       &  532    & 0.14       & 4100    &  9200 & 2.52    & 0.997817 & $-$0.216022 \\
         & 13.0       & 15.3  & $-$38.4         & 11.7 & 21.6           & 13.9& & & & & & & & \\
34       & 23.5       & 1.4   & $-$61.2         & 4.5  & 38.6           & 11.8&  5329       & 2276    & 0.48       & 6000    &  2600 & 3.08    & 0.998001 & $-$0.621582 \\
         & 23.9       & 7.8   & $-$35.3         & 1.4  & 20.4           & 4.0 & & & & & & & & \\
35       & 25.6       & 3.4   & $-$67.5         & 3.9  & 29.0           & 7.8 &  4649       & 2108    & 0.59       & 4700    &  1600 & 5.80    & 0.998375 & $-$0.047991 \\
         & 28.4       & 5.4   & $-$41.3         & 3.4  & 27.0           & 4.4 & & & & & & & & \\
36       & 17.1       & 4.7   & $-$64.7         & 5.0  & 21.9           & 8.2 &  5679       & 1822    & 0.36       & 6000    &  2800 & 5.56    & 0.998327 & $-$0.369874 \\
         & 27.5       & 3.6   & $-$41.7         & 3.5  & 24.1           & 5.7 & & & & & & & & \\
\hline
\end{tabular}
\end{minipage}
\end{table*}

Table~\ref{36.pos} lists central velocities summarised in
Table~\ref{grid.num} in a configuration closer to that shown in
Figure~\ref{grid.spec} so as to illustrate the spatial trend of the
velocity fields. It clearly shows that the single and double peak
regions are segregated. At and to the north-west of the peak of
the H90$\alpha$ RRL emission (Position~13) the single line has the mean
central velocity of $\sim-53$~km~s$^{-1}$. Double peaks almost surround
the above single peak region with the mean central velocities at
$\sim-65$~km~s$^{-1}$ and $\sim-34$~km~s$^{-1}$.

\begin{table*}
\caption{Spatial distribution of central velocities of the H90$\alpha$
RRLs found by the Gaussian profile fit to a spectrum averaged in a
$1.75\times1.75$-arcsec$^{2}$ box (see Figure~\ref{rad.spec}).
Each box lists, from the top to bottom,
the position number as in Figure~\ref{grid.spec}, the central velocity
(velocities if a double peak profile is found at the position) in
km~s$^{-1}$, and non-LTE corrected electron temperature (in K) and
electron density (in cm$^{-3}$).
}
\label{36.pos}
\includegraphics[width=150mm]{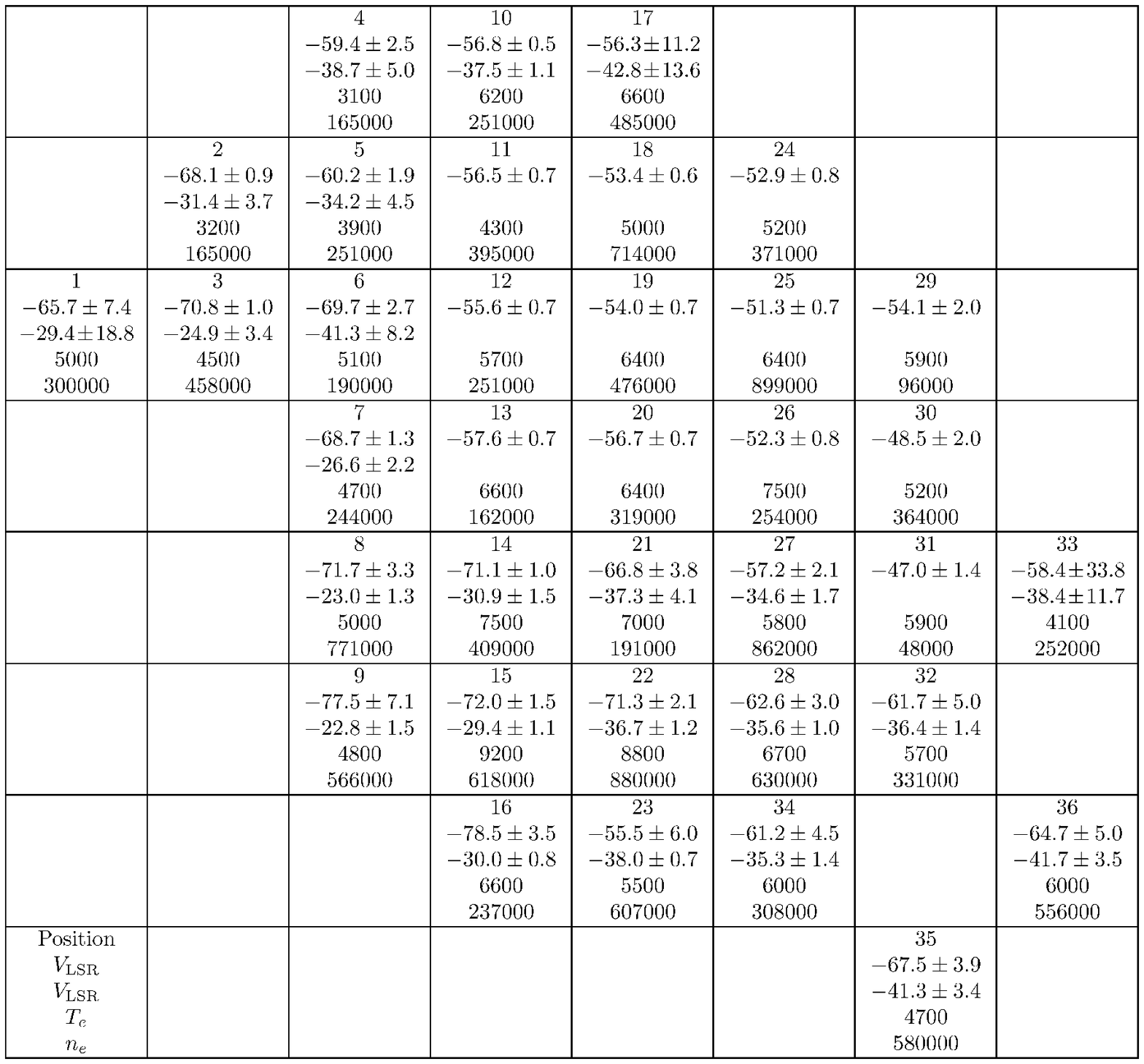}
\end{table*}

\subsubsection{Temperatures}
\label{temps.sect}

The LTE electron temperature is given by (Roelfsema \& Goss 1992)
\[ T_{e}^{*} = \left[ 6943 \cdot \nu^{1.1} \cdot \frac{S_{c}}{S_{l} \cdot
\Delta V_{G}} \cdot \frac{1}{1 + Y^{+}} \right]^{0.87}, \]
where $\nu$ is the frequency in GHz, $S_{c}$ and $S_{l}$ are continuum
and line flux densities in Jy~beam$^{-1}$, respectively, and
$\Delta V_{G}$ is the Gaussian line width in km~s$^{-1}$. Shaver et al.\
(1983) found $Y^{+}$ to be nearly constant in the Galaxy, and we adopted
this value (0.074).

The brightness temperature, $T_{b}$, is given by (e.g.\ Wood \&
Churchwell 1989)
\[ T_{b} = \frac{S_{c} 10^{-47} c^{2}}{2 \nu^{2} k \Omega},\]
where $c$ is the speed of light, $k$ is the Boltzmann's constant, and
$\Omega$ is the beam size. $T_{b}$ is related to $T_{e}$ by
\[ T_{b} = T_{e}(1 - e^{-\tau_{c}}), \]
where $\tau_{c}$ is the continuum optical depth.

$T_{e}^{*}$ may be corrected for non-LTE effects to derive an estimate
of the true electron temperature, $T_{e}$, using (Brown 1987)
\[ T_{e} = T_{e}^{*} \left[b_{n} \left(1-\frac{\beta_{n}\tau_{c}}{2}
\right) \right]^{0.87},\]
where $b_{n}$ and $\beta_{n}$ are the LTE departure coefficients, which
depend on $T_{e}$ and $n_{e}$ (electron density), and can be computed using
the code devised by Brocklehurst \& Salem (1977). At each position, we
calculated $T_{e}^{*}$, $T_{b}$, and $\tau_{c}$. We then created a grid of
$b_{n}$ and $\beta_{n}$ for a range of $T_{e}$ (2000~K $\lid T_{e} \lid$
12000~K, $\Delta T_{e}=100$~K) and $n_{e}$ ($1\times10^{3}$~cm$^{-3}
\lid n_{e} \lid 1\times10^{6}$~cm$^{-3}$,
$\Delta n_{e}=1000$~cm$^{-3}$). Applying these correction factors, we
selected a combination of $b_{n}$ and $\beta_{n}$ (hence $n_{e}$) that
gave the closest corrected $T_{e}$ to the input $T_{e}$. These values
are listed in Table~\ref{grid.num} and in Table~\ref{36.pos} again in a
configuration similar to that shown in Figure~\ref{grid.spec}.

The mean values of $T_{e}$ and $n_{e}$ found in this way are
$\overline{T}_{e}=5700\pm1300$~K and
$\overline{n}_{e}=407000\pm226000$~cm$^{-3}$. The result implies
the fractional
mean-square $T_{e}$ variation [$t^{2} \equiv (\Delta T/T)^{2}$, Ferland
(2001)] for G333.6$-$0.2 is $\sim0.052$. While $t^{2}$ is thought to be
$\sim0$, since a gas in photoionisation equilibrium is nearly isothermal
(Kingdon \& Ferland 1995), relatively high values of $t^{2}$ have been
found in some nebulae (e.g.\ 0.032 in M8, Esteban et al.\ 1999; 0.028 in
Orion, O'dell, Peimbert \& Peimbert 2003). However, it should be noted
that electron temperatures derived from measurements of RRLs, when
corrected for the spatial variation of $Y^{+}$, tend to show smaller
$t^{2}$. For example, Roelfsema, Goss \& Mallik (1992) detected a large
variation of $Y^{+}$ in W3A, and compensating for this fluctuation found
the mean LTE electron temperature to be
$\overline{T}^{*}_{e}=7500\pm750$~K, or $t^{2}=0.01$. We did not
derive $Y^{+}$ but assumed a constant value (0.074, Shaver et al.\ 1983).

\subsection{The Br{\boldmath$\gamma$} image}
\label{brg.sect}

\begin{figure}
\includegraphics[width=84mm]{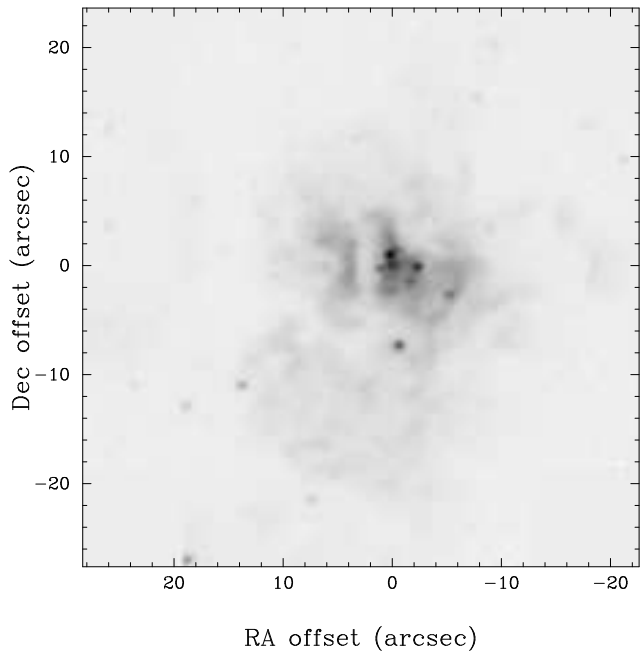}
\caption{A grey-scale image of the 2.17-$\umu$m Br$\gamma$ {\em plus}
the continuum. Coordinates of the origin are as in Figure~\ref{rad.cont}.}
\label{brg.before}
\end{figure}

Figure~\ref{brg.before} shows the 2.17-$\umu$m Br$\gamma$ {\em plus}
the underlying continuum. There are at least 11 point sources
recognisable by eye in the field imaged, and positions of 6 of these
sources coincide with those of NIR sources found common in all the
broadband $JHK'$ frames (Paper~3). It is interesting to note that, as
a simple analysis of relative positions amongst the point sources
reveals, it is not the brightest point source in Figure~\ref{brg.before}
that coincides with the MIR main peak (Paper~1), but the one immediately
below [i.e., the source at offset (0,0)].

\begin{figure}
\includegraphics[width=84mm]{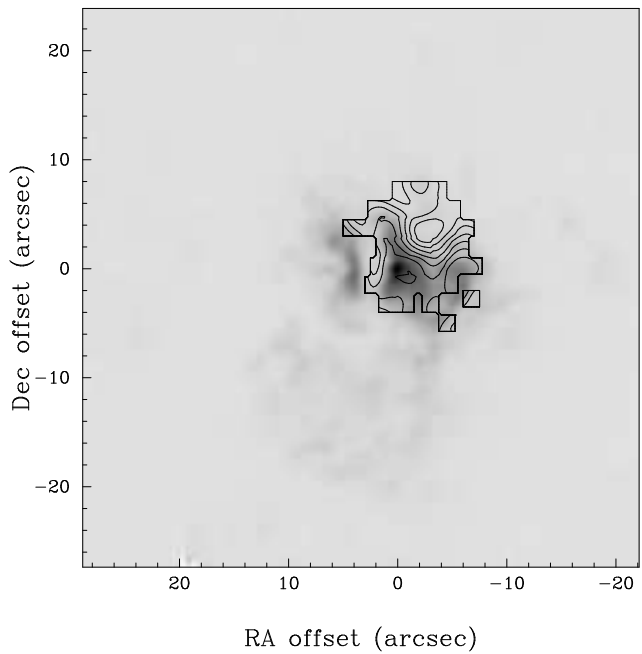}
\caption{The 2.17-$\umu$m Br$\gamma$ image of G333.6$-$0.2. Grey scale:
Same as Figure~\ref{brg.before} but after the
continuum subtraction and a Gaussian smoothing (see Section~\ref{brg.sect}).
Contours: Extinction map (see Section~\ref{extmap.sect}). Contours are
drawn at $A_{V}=4$~mag interval from 12~mag (near the origin) to
36~mag [near ($-$3,4)].}
\label{brg.after}
\end{figure}

Figure~\ref{brg.after} shows the 2.17-$\umu$m Br$\gamma$ image after the
continuum subtraction, which has been Gaussian smoothed to a
slightly lower effective spatial resolution of $\sim0.8$~arcsec.
The morphology seen in Figure~\ref{brg.after} is strikingly similar to
that in the MIR continuum image (see Figure~2, Paper~1). G333.6$-$0.2 has
been found to have a blister geometry viewed face-on (Hyland et al.\ 1980),
and the similar morphology exhibited by the ionised gas and warm dust may be
accounted for if the MIR emitting dust grains are located behind the
H\,{\sc ii} region on the front surface of the photo-dissociation region
(Paper~1). It is also possible that the gas and dust are well mixed but
it is generally thought that dust grains are sublimated at about
1500~K.

Table~\ref{brg.flux} lists the measured Br$\gamma$ fluxes as a
function of the aperture sizes and compares the results of the present
study to the previously published measurements. Each measurement was made
with the aperture centred on the main Br$\gamma$ peak [offset
(0,0)] and the overall uncertainty is estimated to be about 5~per~cent.  
The results obtained are consistent with those from the previous studies,
especially with the measurements of Wynn-Williams et al.\ (1978), and imply
that the continuum subtraction process was reasonably successful.

\begin{table}
\caption{Measured Br$\gamma$ fluxes (in $\times10^{-18}$~W~cm$^{-2}$)
as a function of beam size (arcsec).}
\label{brg.flux}
\begin{tabular}{llllll}
\hline
Beam diameter  & Br$\gamma$ flux                & Br$\gamma$ flux \\
               & [this study]                   & [previous studies] \\
\hline
6.0            & 3.2                            & 2.8$^{a}$ \\
6.5            & 3.6                            & 3.5$^{b}$ \\
30             & 14                             & 7.3$^{a}$ \\
32             & 15                             & 12$^{b}$   \\
\hline
\end{tabular}

\medskip
$^{a}$ Landini et al.\ (1984); $^{b}$ Wynn-Williams et al.\ (1978)
\end{table}

\subsubsection{Extinction map}
\label{extmap.sect}

The theoretical Br$\gamma$ flux (in W~cm$^{-2}$) is related to
the radio flux density $S_{\nu}$ (Jy) measured at frequency $\nu$ (GHz) by
\[ F_{{\rm Br}\gamma} = 8.90\times10^{-6} \alpha_{B} \nu^{0.1}
\Gamma^{-1} S_{\nu} T_{e}^{0.35}, \]
where $\alpha_{B}$ is the total recombination coefficient to the
${\rm n}=2$ level (the `on-the-spot' approximation), and $\Gamma \equiv
\frac{N_{{\rm Lyc}}}{N_{{\rm Br}\gamma}}$ is the number ratio of
Lyman continuum to Br$\gamma$ photons. $\Gamma$ can be computed for a
range of $T_{e}$ and $n_{e}$ using tables of Storey \& Hummer (1995).

The observed flux density is related to the theoretical value in the usual
way, i.e.,
\[ F_{{\rm Br}\gamma}^{{\rm obs}} = F_{{\rm Br}\gamma}^{{\rm theo}}
e^{-\tau_{2.17}}, \]
where superscripts obs and theo represent observed and theoretical fluxes,
respectively, and $\tau_{2.17}$ is the optical depth at the wavelength of
the Br$\gamma$ emission. The extinction at a given wavelength $\lambda$ is
related to the optical depth by
\[ A_{\lambda} = 2.5\tau_{\lambda}\log_{10}e.\]
Finally, assuming the extinction at 2.17~$\umu$m, $A_{2.17} \simeq A_{K}$,
the extinction found at 2.17~$\umu$m can be extrapolated to the visual
extinction at 0.55~$\umu$m using the NIR extinction law derived by
Martin \& Whittet (1990)
\[ A_{K} \simeq 0.09 A_{V}. \]

For $T_{e}$, the nearest to the mean value derived earlier (5700~K)
in the table of Storey \& Hummer (1995), 5000~K was used.
The mean value of $n_{e}$ obtained at the same time showed quite a large
standard deviation ($\overline{n}_{e}=407000\pm226000$~cm$^{-3}$);
however, the 2.17-$\umu$m Br$\gamma$ emission line is not very sensitive to
electron densities in the range usually found in H\,{\sc ii} regions.
Populations of levels are more susceptible to density in high n
states, such as those found at radio frequencies, as incoming free
electrons have a higher probability of perturbing the bound electrons
into a different orbit. Whereas low n transitions [such as Br$\gamma$
(${\rm n}=$ 7$-$4)] are affected less by collisions. We nonetheless
used a mean value in the density range $n_{e}=10^{5}-10^{6}$~cm$^{-3}$
in the calculation of the theoretical Br$\gamma$ flux.

Some positions exhibit a moderate optical thickness at 3.4~cm (see
Table~\ref{grid.num}). $S_{\nu}$ at each position has therefore been
multiplied by a correction factor, [$\tau_{c}/(1 - e^{-\tau_{c}})$],
which varied from 1.03 at one end of the region mapped with a reasonable
$S/N$ (Position~17) to 1.64 at the peak (Position~13). However,
the correction can only be made in regions where $\tau_{c}$ is
available, i.e., only in the 36 $1.75\times1.75$-arcsec$^{2}$ boxes.
This will confine the extent of the extinction map to the central
$\sim10\times10$-arcsec$^{2}$.

Landini et al.\ (1984) measured $A_{V}=18.5$~mag in a 6-arcsec
aperture centred on the $K$-band peak. As discussed in
Section~\ref{missing.flux}, radiation from extended structures escaped
detection in our current radio interferometric observations. We assumed
that the distribution of the missing flux underlying the central compact
structure is a 2-D Gaussian. We then adjusted its peak value and FWHMs so
that it would contain the amount of the missing flux and contribute
sufficient emission in the centre to yield the level of the Landini
et al.'s extinction measurement. This was achieved by adding a 2-D
Gaussian with a peak value of 0.4~Jy beam$^{-1}$ and FWHMs of
20.5~arcsec; however, we note that recently released {\em Spitzer}
GLIMPSE (Benjamin et al.\ 2003) images ({\em IRAC} 3.6, 4.5, 5.8, \&
8.0~$\umu$m) of the G333.6$-$0.2 region show a wide spread
($\sim2\times2$~arcmin$^{2}$), fairly uniform faint emission around
more intense (in fact, saturated) central object. Therefore, it is
quite possible that there is a large scale pedestal-like extended
component to the missing flux, in addition to the 2-D Gaussian structure.
If so, the peak value and FWHMs we used for the Gaussian missing flux
component are likely to be their respective upper and lower limits.
Consequently, our map is more suited to demonstrate the spatial
trend in the extinction rather than to show the exact extinction values.
It is also noted that, although altering the peak value and FWHMs of
the 2-D Gaussian missing flux component changed the finer details of the
resultant extinction map, the overall spatial trend remained virtually
the same in the central region. Even adding a uniform missing flux
produced a very similar trend.

Figure~\ref{brg.after} shows the extinction map (contours) superimposed
on the Br$\gamma$ image of G333.6$-$0.2. Lower extinction values
are found near the Br$\gamma$ peak [offset (0,0)] and in the
western extension [around offset ($-$6,$-$1)], i.e., near and around
the ionising sources. Point sources with the brightest apparent $K$
magnitudes and the reddest $H - K$ colours were found at these
positions, and the large $H - K$ colour excess of those objects
most probably indicates the physical association of the point
sources with the H\,{\sc ii} region (Paper~3). This trend seems to
fit with the blister nature of G333.6$-$0.2.

A considerable number of H\,{\sc ii} regions have been found to have a
blister geometry in which the H\,{\sc ii} regions are located at the
edges of molecular clouds (Habing \& Israel 1979). The blister geometry
is thought to be the result of the champagne phase in which H\,{\sc ii}
regions expand preferentially toward directions of decreasing density
(e.g.\ Yorke 1986). The blister H\,{\sc ii} regions are therefore
ionisation bounded on the molecular cloud side and density bounded on
the side of outward champagne flow (Yorke, Tenorio-Tagle \& Bodenheimer
1983). The density of ionized gas decreases approximately exponentially
away from the cloud and a large fraction of ionizing photons can escape
from the density bounded side. That is, that side can be relatively
optically thin. Hyland et al.\ (1980) found that G333.6$-$0.2 is a blister
H\,{\sc ii} region viewed face-on. The density bounded side then is the
front face and the low extinction values found around the ionizing sources
are perhaps natural consequences of this geometry.

Taken at face value, the presence of red objects in lower extinction
regions may seem contradictory; however, it was shown in Paper~3 that,
in the colour-colour diagram space, for example, these red sources
are found significantly away from the average interstellar extinction
curves. It was also demonstrated that emission from hot ($\gtrsim600$~K)
dust grains, which are likely to be located in circumstellar discs, is
considerably more important in the $K$-band
than at $H$, resulting in the large $H-K$ colours of these objects.
In other words, the red colours are due not to the extinction but to the
intrinsic infrared excess, and so it is not implausible for these sources
to be found in lower extinction regions.

There is a slight increase in extinction between the main and
secondary peaks. In a clumpy medium, an expansion front (weak ionisation
front with its associated shock) of an H\,{\sc ii} region passes around
the more dense clumps of neutral material and compresses them (Yorke 1986),
which results in embedded neutral globules and dark lanes sometimes seen
in optical images of emission nebulae. As mentioned earlier, there is likely
to be an ionizing source both in the main and secondary peaks and the gap
between the two peaks observed at NIR and MIR wavelengths could be a
pressure compressed dense region that consequently shows a high extinction,
rather than a foreground extinction lane. The fact that this gap is also
observed at radio frequencies (see Figures~\ref{rad.cont} \& \ref{rad.line})
may support this suggestion.

The highest extinction is located near offset
($-$3,4). It was suggested in Paper~2 that there is likely to be an
extra overlapping of dense, cold material containing silicates at
5~arcsec north of the main MIR peak, since the spectrum taken at
the position showed a deep silicate absorption. The beam size used for
the 5$.\!\!''$0N position was 5.6~arcsec in diameter and most likely
included this region of high extinction. The position coincides with
a radio peak (see Section~\ref{vel.revisit}) and the high extinction
may be associated with this radio source.

\subsubsection{The Lyman continuum photons}

The theoretical Br$\gamma$ flux can also be expressed as
\[ F_{{\rm Br}\gamma}^{{\rm theo}} = \frac{N_{{\rm Br}\gamma}}{4 \pi D^{2}} \ \frac{hc}{\lambda_{{\rm Br}\gamma}}, \]
where $D$ is the distance to the object (in cm), $h$ is the Planck's
constant, and $\lambda_{{\rm Br}\gamma}$ is the wavelength of the Br$\gamma$
emission. And recalling that
$\Gamma \equiv \frac{N_{{\rm Lyc}}}{N_{{\rm Br}\gamma}}$,
\[ N_{{\rm Lyc}} = 4 \pi D^{2} \frac{\lambda_{{\rm Br}\gamma}}{hc} F_{{\rm Br}\gamma}^{{\rm theo}} \Gamma. \]
Again by knowing $S_{\nu}$, $n_{e}$ and $T_{e}$, and using an appropriate
table from Storey \& Hummer (1995), the number of Lyman continuum photons
can be estimated.

In constructing the extinction map, we have assumed a 2-D
Gaussian-shaped missing flux distribution. This also means
that the missing flux is concentrated and that
$\sim100$~per~cent of the total flux at 3.4~cm (80~Jy, McGee et al.\
1975) comes from the inner region. However, such a high concentration
of flux in the centre is most probably the upper limit as there appears
to be a wider spread of faint emission further out (see above). The
number of Lyman continuum photons in the inner $\sim50$-arcsec is thus
estimated to be $\lesssim9.5\times10^{49}$~s$^{-1}$.

Rubin et al.\ (1994) suggested that
the effective temperature ($T_{{\rm eff}}$) of star(s) exciting
G333.6$-$0.2 is $\sim36000$~K which, according to the stellar
spectral-type scale of Martins, Schaerer \& Hillier (2005), is
approximately characteristic of an O7V star. The number of the Lyman
continuum photons emitted by an O7V star is estimated to be
$\sim5\times10^{48}$~s$^{-1}$ (Martins et al.\ 2005). Therefore,
if all the exciting stars in G333.6$-$0.2 were of the same spectral type,
there would be up to 19 O7V stars.
In Paper~1, it was found
that about a dozen O8V stars ($T_{{\rm eff}}=36000$~K) is required to
radiate the ionising photon flux necessary to produce the observed
[Ne\,{\sc ii}] central dip.\footnote{Note that the stellar parameter
calibration of Martins et al.\ (2005) takes into account effects of
line-blanketing which shifts (lowers) the effective temperature for
each stellar spectral type. However, for a given stellar effective
temperature, other stellar parameters (e.g.\ luminosity, etc.) are
essentially unchanged.}

\subsection{Velocity structure (revisited)}
\label{vel.revisit}

The double peak of an RRL observed in an H\,{\sc ii} region could be
interpreted as:
\begin{description}
\item (a) expansion of shell-like structure,
\item (b) rotation,
\item (c) two H\,{\sc ii} regions at different velocities, or,
\item (d) outflows/jets.
\end{description}
We discuss each of these in turn, followed by a short section (e),
describing details of a toy model for our preferred option, (d).

(a) expansion of shell-like structure. There is no obvious shell-like
morphology seen in the regions where double peaks are observed. The mean
velocity difference between the two lines is $\sim30$~km~s$^{-1}$.
For a shell-like structure to be unresolved ($<1.6$~arcsec or
$<0.023$~pc) and to have an expansion velocity of $\sim30$~km~s$^{-1}$
would require a dynamical age of only a few hundred years. This seems
rather unlikely. Also, the stellar-wind blown cavity scenario for the
[Ne\,{\sc ii}] central dip was ruled out in Paper~1, since there were
no signs of such a cavity in the continuum image at a nearby wavelength.

(b) rotation. This appears unlikely since there is no systematic shift
of central velocities across the region to suggest a rotation. 

(c) two H\,{\sc ii} regions at different velocities. Combined with the
multiple point sources seen in the NIR images (see
Figure~\ref{brg.before} and Paper~3), it could be that the double peak
profiles of the H90$\alpha$ RRL in G333.6$-$0.2 represent the existence
of several compact H\,{\sc ii} regions at different velocities.
In the $JHK'$ images (Paper~3), all of which were taken at a similar
spatial resolution to that of the radio maps presented here, at least
four point sources were identified in the central region of G333.6$-$0.2,
two of which are found in the main peak (cf.\ NM and SM in Figure~7,
Paper~3), one in the secondary peak to the east of the main, and one in
the south-west extension. Double peak regions do indeed coincide with the
MIR main peak and the south-west extension. At the highest spatial
resolution so far attained ($\sim0.6$~arcsec), the 2.17-$\umu$m image
(Figure~\ref{brg.before}) reveals even more star-like point sources in
the central region and there appears to be three sources located very
close together near the position of the MIR main peak.

\begin{figure*}
\centering
\includegraphics[width=150mm]{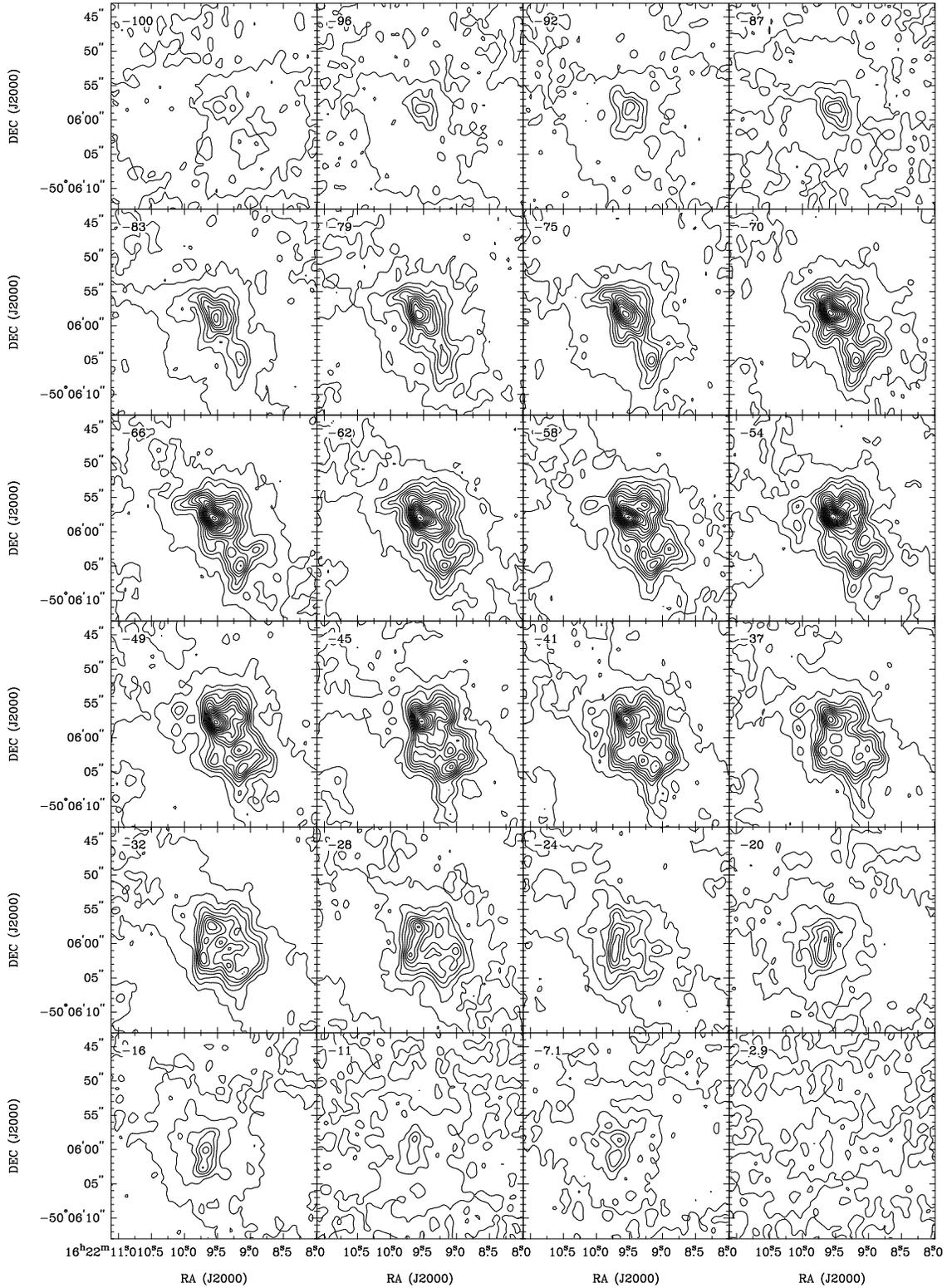}%}%
\caption{The H90$\alpha$ RRL maps in each velocity plane
($\sim4$~km~s$^{-1}$ wide). The velocity plane is indicated at the top
left hand corner of each plot. The contours are drawn at a 5~per~cent
interval from 0 to 100~per~cent of the maximum flux density in each
plane. The maximum values in each plane are 1.4, 1.8, 2.2, 2.3, 3.8,
5.0, 5.8, 8.1, 8.0, 8.3, 8.4, 9.2, 7.5, 6.9, 6.0, 5.3, 4.0, 3.7, 3.2,
2.8, 2.5, 1.5, 1.9, and 0.82 $\times10^{-2}$~Jy~beam$^{-1}$ from $-100$
to $-2.9$~km~s$^{-1}$.}
\label{vel:figs}% label for figure
\end{figure*}

Figure~\ref{vel:figs} shows intensity distribution in each velocity
plane ($\sim4$~km~s$^{-1}$ wide) in the range $\sim-100$ to
$0$~km~s$^{-1}$. This sequence of velocity maps also reveals the
multiple nature of the exciting source in the central region of
G333.6$-$0.2. The main radio peak dominates throughout the velocity
range because of its large line width ($\sim50$~km~s$^{-1}$ FWHM).
The ridge to the south of the main peak is visible up to the
$-62$~km~s$^{-1}$ frame but is absent till another peak appears further
south at about $-50$~km~s$^{-1}$. The south-west extension seems to
contain at least three peaks (see also Figures~\ref{rad.cont} \&
\ref{rad.line}). As mentioned earlier, there is at least
one point source identified in the NIR images in the south-west
extension (see Figure~\ref{brg.before} and Paper~3).

The shape of the radio main peak is rather complex. There appears to be
another peak close (about 2~arcsec north-west) to the main peak
(see the $-$58~km~s$^{-1}$ frame). The extinction map (Figure~\ref{brg.after})
shows a large increase in the extinction value around this position,
which may imply that obscuring dust grains are associated with this
peak. The main peak and this nearby local maximum do not have obvious IR
counterparts and could be younger than the other exciting sources in
the H\,{\sc ii} region, which in turn may also explain the presence
of heavy obscuring dust grains. If this supposition of the difference
in age were correct then this alone would lead to the conclusion of
multiple exciting sources, and hence the presence of several compact
H\,{\sc ii} regions, in G333.6$-$0.2.

If flat Galactic rotation curves, with a constant circular velocity
$\Theta=220$~km~s$^{-1}$ and the Sun's distance from the Galactic
centre $R_{0}=8.5$~kpc, were adopted, the difference in velocity of
$\sim30$~km~s$^{-1}$ between two peaks in the double peak regions would
imply a separation of the H\,{\sc ii} regions along the line of sight
(los) of $\sim1.6$~kpc. While possible, it seems rather unlikely that
several unassociated compact H\,{\sc ii} regions just happen to coincide
along the los to produce the velocity structure observed. Of course,
G333.6$-$0.2 is an active star-forming complex and adopting the Galactic
rotation curves to interpret the velocity structure may not be appropriate.

\begin{figure}
\includegraphics[angle=270,width=84mm]{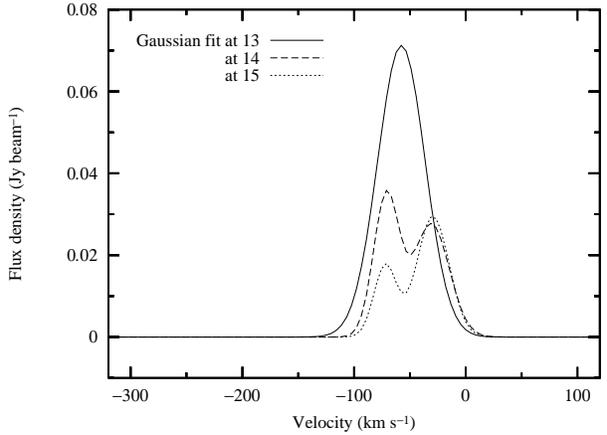}
\caption{Gaussian profiles at Positions~13, 14, and 15 (see also
Figure~\ref{rad.spec}).
}
\label{peak.ridge}
\end{figure}

Another interesting feature that is difficult to interpret with the los
coincidence scenario for the complex velocity structure presents itself
when the single and double peak spectra are compared closely.
Figure~\ref{peak.ridge} shows the
best-fit Gaussian profiles at Positions~13, 14, and 15.
It is remarkable that
the overall shape of the spectra matches almost exactly and the double peak
appears more or less symmetric with the single peak between the two peaks.
Again it seems rather unsatisfactory to explain such features as a chance
coincidence of compact H\,{\sc ii} regions along the los. 

The spatial resolution of the radio maps (1.6~arcsec), despite being the
highest so far obtained at these frequencies and for the first time
revealing the double peak profile of the H90$\alpha$ RRL, is more than
twice the highest resolution of the IR images ($\sim0.6$~arcsec). To
examine more closely the relationship between the IR point sources and
the velocity structure, radio interferometric observations at even
higher spatial resolution, at least that comparable to the IR images,
are required. IR imaging both at high spectral and spatial resolution is
perhaps an alternative.

(d) outflow or jets. There are no obvious structures that suggest
outflows or jets in the double peak regions. However, it is possible
that outflows/jets are inclined in the los so that such structures are not
immediately obvious, or are unresolved by the 1.6-arcsec beam. Yorke et
al.\ (1983) produced theoretical radio continuum maps of H\,{\sc ii}
regions in the champagne phase, which occurs when the ionisation front
of an H\,{\sc ii} region encounters a region of strong density gradients,
such as the boundary of a molecular cloud. They found that the champagne
outflow (or preferential expansion of an H\,{\sc ii} region along negative
density gradients) can attain high velocity ($>10$~km~s$^{-1}$) and in
the case of a slab-like geometry it can be bipolar. They also found that
in general the maximum of radio emission, which is proportional to
$n_{e}^{2}$ (e.g.\ Rubin 1968), does not coincide with the ionizing source.
This may explain the offset between the radio and IR sources.

The champagne outflow scenario may also explain the broad line width
of the single peak. The outflows caused by the preferential expansion of
the H\,{\sc ii} region into less dense regions may not be well-collimated.
In fact, the velocity structure depicted in the models of Yorke et al.\
(1983) shows a fan-like radial outpouring of ionized gas from the mouth
of the outflow. If an RRL were observed from such a structure its spectrum
would probably show a large line width due to a range of los velocities
present in the radial velocity field.

It was found in Paper~2 that the intrinsic magnetic field lines follow
well the flux distribution north-west of the MIR main peak. This
morphology is most likely to be due to the presence of a clump of dense
material containing silicates (indicated by the shape of the spectrum
taken at 5~arcsec north of the MIR main peak). There is a local visual
extinction maximum about 4~arcsec north and 3~arcsec west of the IR main
peak (Figure~\ref{brg.after}), which may be associated with the radio peak
$\sim2$~arcsec north and $\sim3$~arcsec west of the radio main peak seen
in the $-58$~km~s$^{-1}$ plane (see Figure~\ref{vel:figs}). The expanding
H\,{\sc ii} region may have carried the frozen-in, threading magnetic
field with its expansion. If such structure then encountered the
aforementioned clump of dust and compressed the magnetic field against it
the intrinsic polarization pattern may result. Near the boundary, where the
expansion hit the clump, the radio continuum peak may have been produced
by the high density.

\begin{figure}
\includegraphics[width=84mm]{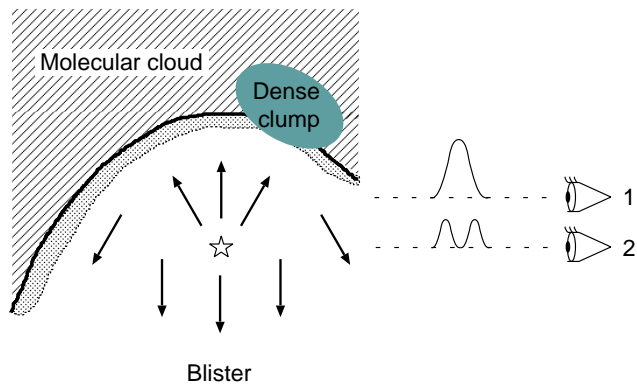}
\caption{A `cartoon' representation (a side view) of a possible
situation in G333.6$-$0.2. \ding{73} denotes the IR exciting source(s).
Two different lines of sight (1 and 2) are also indicated, together with
a probable RRL profile along each los. Arrows are drawn only to illustrate
likely direction of the champagne flows. See text for details.}
\label{poss.sit}
\end{figure}

(e) a toy model. In view of the ability of the outflow model to account
for the general appearance, we present (Figure~\ref{poss.sit}) a more
elaborate toy model of G333.6$-$0.2, which can explain all of the
key features. The blister geometry, that may account for the trend
found in the extinction map (Figure~\ref{brg.after}), may not be exactly
face-on but somewhat tilted south-east. Towards the north of the exciting
(IR) source(s), as we look through a range of los velocities in the
fanned-out champagne outflow in the relatively high density regions near
the molecular cloud wall, we observe the strong, broad single RRL (los~1,
Figure~\ref{poss.sit}). We obtain double RRL spectra towards the south-east,
as faster flows are found along the wide opening of the blister `bowl',
whilst flows in the intermediate velocity range are generally fainter
(as the region is more diffuse) and relatively perpendicular with respect
to the los (los~2, Figure~\ref{poss.sit}). Champagne flow velocities are
expected to be greater further away from the molecular cloud (Bedijn \&
Tenorio-Tagle 1981) and we are possibly observing this phenomenon as there
seems to be an apparent trend that the blue peak shifts to the bluer
velocity (and the red redder) the more south we look (see
Table~\ref{36.pos}).

Another interesting feature, which may be explained by this picture, is that
in the north the blue peak is stronger, whereas the red peak becomes
brighter in the south (see Figures~\ref{grid.spec}, \ref{rad.spec},
and \ref{peak.ridge}). This could be due to the presence of the dense clump.
The density near the front side of the molecular cloud wall is enhanced by
the dense clump and therefore, near the clump, we find strongest emission
from the flows that are directed towards us; south of the clump, emission
from the back wall (away from us) dominates. Note that, whilst the strength
of the blue peak changes dramatically depending on the exact position, the
red peak brightness remains more or less constant.

\section{Conclusions}

The radio interferometric observations and NIR imaging of the southern
massive star-forming region G333.6$-$0.2 are presented. The morphologies
of the high spatial resolution (1.6-arcsec) 3.4-cm continuum and
H90$\alpha$ RRL maps are remarkably similar to each other and to the IR
images. However, the radio main peak neither coincides with the IR main
peak nor has any obvious IR counterparts. The H90$\alpha$ RRL spectra
show a complex structure with a double peak profile at some positions
observed for the first time. The complex velocity structure may
be explained by champagne outflows, which may also explain the offset
between the radio and IR sources. The 3.4-cm radio continuum map is
combined with the 2.17-$\umu$m Br$\gamma$ image in constructing the
visual extinction map. The blister geometry of the H\,{\sc ii} region
probably causes the trend seen in the extinction map. The number of
Lyman continuum photons in the central 50-arcsec is estimated to be
equivalent to that emitted by $\lesssim19$ O7V stars.

\section*{Acknowledgments}

We would like to thank the respective staff at ATCA and AAT for their
assistance during our observing runs. We acknowledge with thanks
the MIRIAD help given by Dr.\ Neil Killeen.
We also thank the anonymous referee for valuable comments.

\label{lastpage}

\end{document}